\documentclass[seceq]{ptptex}
\usepackage{wrapft}

\usepackage{graphicx}

\title{
A brief review of long-term X-ray and optical variability in radio-quiet AGN
}

\author{
Philip \textsc{Uttley}$^{1,2}$\footnote{E-mail: pu@milkyway.gsfc.nasa.gov} 
and Ian M. \textsc{M$^{\rm c}$Hardy}$^{1}$}

\inst{
$^1$School of Physics and Astronomy, University of Southampton, Southampton SO17~1BJ, UK\\
$^2$Laboratory for High Energy Astrophysics, Code 662, NASA Goddard Space Flight Center, Greenbelt, MD 20771, USA
}

\abst{
Long-time-scale X-ray and optical variability is a key characteristic of AGN.
Here, we summarise our current understanding of the X-ray and optical
continuum variability of radio-quiet AGN
and the relation between the two bands. We demonstrate the
strong connection between the X-ray variability properties of AGN and the
variability of stellar-mass black hole candidates on much shorter time-scales,
and discuss the implications of this result for the origins of the
variability.  The relationship between optical and X-ray
variability is complex, with some AGN showing strong X-ray/optical
correlations while others show no obvious correlation. We suggest a 
possible explanation for this variety of behaviour.
}

\begin{document}

\maketitle

\section{Introduction}
Time-variable emission over the entire observable spectrum is one of the 
defining characteristics of AGN.  Variability on time-scales of months to 
years provided the first key evidence that the emitting regions were 
extremely compact, leading to the suggestion that AGN are powered by 
massive black holes.  However, although the black hole paradigm has 
grown stronger due to a variety of subsequent observations, the origin
of the variability largely remains a mystery.  In radio-loud AGN, 
some progress has been made in understanding the broadband variability in 
terms of jet models of emission\cite{mch99}, but the situation is less clear in
radio quiet AGN, which form the bulk of the AGN population. 
Because, in the optical
waveband, the variability is fairly slow, it can only be studied in detail 
with long, well-sampled monitoring campaigns which are difficult
to organise.  In the X-ray band, where the variability is much more rapid,
short-term variability was originally
studied using `long-looks' of a day or more duration, by X-ray
satellites such as {\it EXOSAT} and {\it ASCA}, but longer time-scales
were inaccessible due to the constraints of scheduling and pointing these 
satellites.  In 1995, the launch of the Rossi X-ray Timing Explorer 
({\it RXTE})
revolutionised the study of AGN variability, because the rapid slewing 
capability and flexible scheduling of {\it RXTE} allowed well-sampled 
long-term monitoring of AGN X-ray variability for the very first time. \\ 
With {\it RXTE}, it has been possible to study X-ray variability of
radio-quiet AGN over a very
broad range of time-scales for comparison with the (as it turns out)
remarkably similar variability properties of stellar mass black holes in 
X-ray binary systems (BHXRBs).  Also, it has been possible
to compare the long-term
X-ray variability with the optical variability sampled by a few optical 
monitoring programs, to examine the relationship between the two bands,
which we might expect to be dominated by different emission mechanisms
(optically thin versus optically thick).  In this paper we will review 
our current understanding of the long-term X-ray variability of radio-quiet
AGN, and how it relates to the X-ray variability on shorter time-scales. We
will also consider the relationship between the X-ray and optical bands, and
discuss models which might explain the variability in both bands.

\section{X-ray variability and the AGN-BHXRB connection}
\subsection{Before RXTE}
In the 1980s, studies with {\it EXOSAT} showed that on short time-scales,
AGN variability appeared to be red-noise.\cite{mch87,law87
} \ In other words the variability 
showed no obvious periodic or quasi-periodic behaviour (and hence is 
called noise),
but showed variability over the entire range of sampled time-scales 
(and hence is
`red' with the variability power density
increasing towards lower temporal frequencies).
{\footnote{We note here that to date, no statistically significant
examples of (quasi)-periodicities have been found in AGN X-ray light 
curves\cite{ben01} with the possible (and intriguing) exception of long {\it EUVE} observations of two Narrow Line Seyfert~1s.\cite{hal03} \ This 
situation is probably to be expected, because 
assuming such signals exist in AGN
at equivalent frequencies (i.e. scaled by black hole mass)
and powers to those seen in BHXRBs, only very long
(weeks) and continuous observations would be likely to detect them.}}  The 
corresponding power spectral density functions (PSDs) were
described by power-laws (of index -1 to -2), 
which were unbroken down to the lowest sampled \
frequencies of $\sim10^{-5}$~Hz.  However, it was 
noted\cite{mch88} that this power-law shaped PSD is reminiscent of the high
frequency PSDs (above $\sim1$Hz) of black hole X-ray binary systems, 
and that if the similarity holds to lower frequencies we should expect to
see a break to a flatter PSD slope, below frequencies corresponding
to time-scales of days-weeks (provided we make the 
physically plausible assumption that the break time-scale 
scales linearly with black hole mass).
Attempts to detect this break frequency,
using sparsely sampled archival data from different missions to cover
long time-scales, were hampered by the distorting effects of sampling and
the limited data, but they were at least suggestive that there was a
flattening of the PSD at the expected frequencies.\cite{mch88,pap95} \\
Around the same time as the early AGN PSD studies,
an inverse correlation was noted between the amplitude of variability in 
$\sim$day-long AGN X-ray light curves and the X-ray luminosity of the 
AGN.\cite{bar86,gre93,nan97} \ Of a number of possible models to explain this 
result, the most promising was that the effect represented the expected
correlation
between the black hole mass (which is tracked by luminosity assuming a common
fractional accretion rate) and the variability time-scale, with more massive
(and luminous) AGN showing slower variability due to their larger size.
The situation was complicated by the discovery that Narrow Line Seyfert~1s 
(NLS1) do not follow this correlation (they show large 
variability amplitudes even at high luminosities).\cite{tur99,lei99}

\subsection{The RXTE era}
\begin{figure}[htb]
 \parbox{\halftext}{
\centerline{\includegraphics[width=6cm,height=6.6cm,angle=-90]{uf3516_4051.ps}}
                \caption{Comparison of PSDs of NGC~4051 and NGC~3516 with
typical PSDs observed for Cyg~X-1.  The PSD of NGC~4051 is constructed from a combination of {\it XMM-Newton} high-frequency data and long-term {\it RXTE} monitoring\cite{mch04}.}\label{nfnpsd}}
 \hfill
 \parbox{\halftext}
{\centerline{\includegraphics[width=6.6cm,height=6cm]{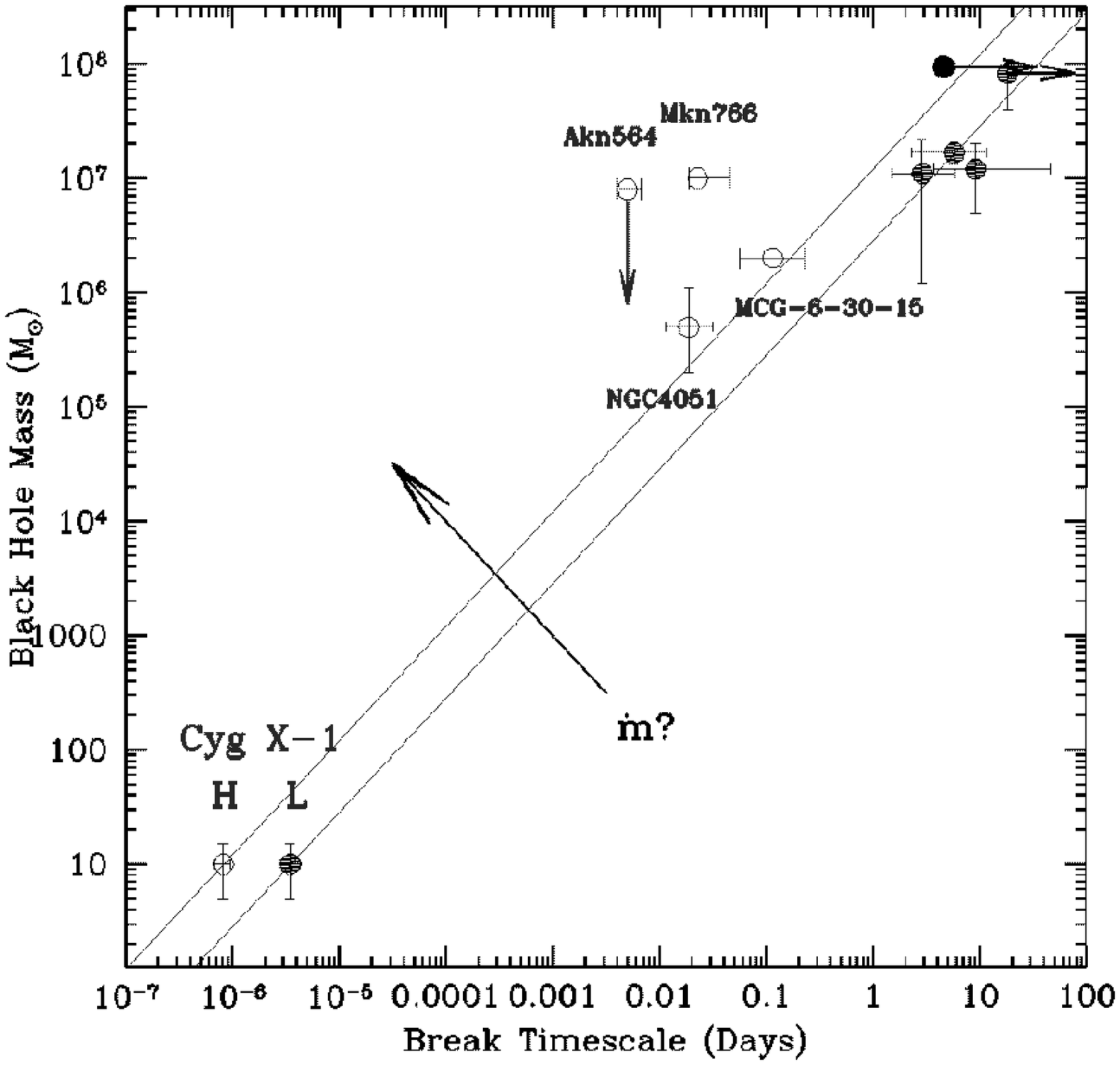}}
                \caption{Comparison of PSD break time-scale with black hole
 mass.  BH masses are estimated using either optical
reverberation mapping or host galaxy bulge properties.  AGN marked with 
open circles correspond to NLS~1 or borderline NLS1 cases 
(e.g. MCG--6-30-15, which is not classified as NLS1 but often described 
as such).  See elsewhere for further details.\cite{mch04}}\label{masstime}}
\end{figure}

The prospect of a direct analogy between AGN and BHXRB X-ray variability 
led a number of researchers\cite{mch88,hay98,cze01}
to show how the X-ray PSD might be used
to estimate the black hole mass of the AGN, by simply scaling from the
PSDs of BHXRBs (for example the well-studied Cyg~X-1) and assuming a
BHXRB black hole mass of $\sim10$~M$_{\odot}$.  However, without actual
confirmation of the PSD break time-scales in AGN, it was impossible to tell if
the variability time-scales really did scale with black hole mass, or indeed
if AGN variability really is similar to that of BHXRBs{\footnote{It is
worth noting
here than BHXRBs show a wide variety of variability properties, with
different PSD shapes (and energy spectra) 
depending on the `state' of the BHXRB.\cite{mcc03} \ For example, the
low/hard state PSD is characterised by two breaks, a high frequency
break (where PSD index changes from $\sim-2$ to -1) 
around 1-6~Hz, and a low frequency break around $\sim0.1$~Hz (where PSD 
index changes from -1 to 0).  The high/soft state PSD on the other hand shows
only a single break (from index $\sim-2$ to -1) at around 10~Hz.  Therefore, 
when estimating black hole mass using PSDs
it is important to know which BHXRB state (if any) we are comparing with.}}.
\begin{wrapfigure}{r}{6.6cm}
\centerline{\includegraphics[width=6.6cm,height=7.5cm]{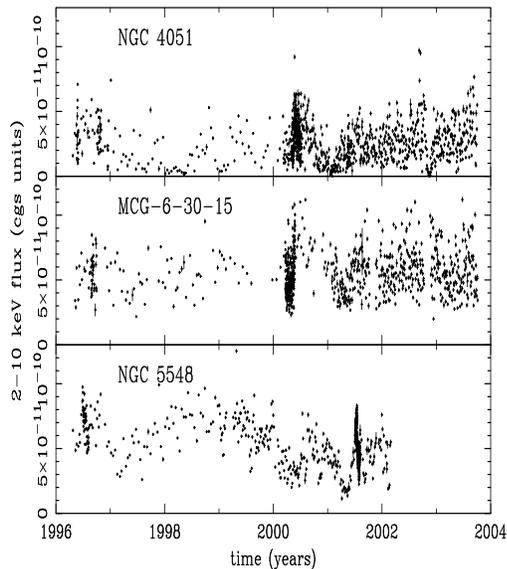}} 
\caption{Long-term 2-10 keV X-ray monitoring light curves of 3 Seyfert galaxies.}
\label{xtelcs}
\end{wrapfigure}
The situation changed with the launch of {\it RXTE} and the first 
high-quality AGN X-ray monitoring campaigns (see Fig.~\ref{xtelcs}),
which sampled a broad range of time-scales.  One notable aspect
of the long-time-scale data is that, although the variability amplitudes
of AGN with different luminosities are very different on short time-scales,
they are similar on long time-scales
(e.g. compare NGC~4051 with the 100 times more luminous 
NGC~5548 in Fig.~\ref{xtelcs}), as would be 
expected if there is a PSD break at a time-scale which scales with 
the mass.\cite{mar01} \
Within a few years, sufficient
long-term monitoring data was available to confirm the existence of 
PSD breaks. \cite{mch98,ede99,utt02,mar03,pap03} \ 
Detailed analyses, using Monte Carlo methods to robustly constrain
the PSD shape (e.g. accounting for 
`aliasing' effects), were able to show that the PSD appeared to
break from an index$\sim-2$ to index$\sim-1$ and not 0, 
implying that the breaks are more likely to be
analogous to the high-frequency breaks in
the PSD of Cyg X-1 (in either the low or high state), and that the 
break time-scales were consistent with a linear scaling of characteristic
time-scales with black hole mass.\cite{utt02,mar03} \\
Several AGN show PSD breaks at high enough frequencies that they
are detectable in {\it XMM-Newton} long-look
observations.\cite{vau03a,vau03b,mch04}\ The time-scales of
these breaks are
too short for these AGN to be analogues of low-state BHXRBs, since
the inferred low masses (estimated by scaling from the low-state
PSD break in Cyg X-1)
would imply super-Eddington accretion rates, 
inconsistent with the low accretion rate thought to be associated with 
the low state. One possible explanation is that 
these AGN show PSDs which are analogous to the PSDs of BHXRBs in the high/soft
state, which show higher break frequencies than in the low/hard state.
However,
the best way to distinguish the low and high-state PSD shapes is to look at
even lower frequencies where - in the low state -
we expect to see a second break to zero slope,
about a decade below the high-frequency break.  
The best quality AGN PSD
yet obtained is that of the NLS1 NGC~4051\cite{mch04}, which is 
reproduced here in Fig.~\ref{nfnpsd}, plotted for comparison with Cyg~X-1
as frequency$\times$power\cite{hay98}so 
that a flat top corresponds to an index of -1.  Clearly there is no 
low-frequency break and the PSD looks much more similar to that of 
Cyg~X-1 in the high/soft state. \\
By scaling the PSD break time-scale seen for NGC~4051 with the equivalent
break time-scale in Cyg~X-1, we 
estimate a low black hole mass of $3\times10^{5}$~M$_{\odot}$, consistent with
that obtained by reverberation mapping.\cite{she03} \
Fig.~\ref{nfnpsd} also shows the PSD of the broad-line Seyfert
NGC~3516, which shows a break at longer time-scales, as one would expect given the larger black hole mass of this
AGN from reverberation mapping.
Because of the lower break frequency the
PSD frequency coverage
does not extend far enough to rule out a low/hard state PSD in this case,
however there is tantalising evidence of a low state PSD in another 
broad line Seyfert, NGC~3783.\cite{mar03} \
Interestingly, a comparison of break time-scales with
black hole masses estimated by various means (Fig.~\ref{masstime})
is suggestive that the
NLS1s tend to have shorter time-scales for their mass, which are consistent
with high/soft state PSDs, or more generally implies that there is
a decrease in variability
time-scale with increasing accretion rate.\cite{mch04}
If we assume that 
broad line Seyferts have different accretion rates as a class to NLS1s,
the difference in PSD break time-scales could help explain why NLS1s do not
conform to the variability amplitude-luminosity correlation observed in broad line Seyferts.\cite{pap04} \

\subsection{Physical implications}
The fact that the
broadband PSDs of AGN measured by {\it RXTE} can be described as singly
broken or more gently bending continua\cite{mch04} suggests 
that the long-term variability is a continuation of the same 
red-noise process seen on shorter time-scales.  More importantly, the fact that
AGN X-ray variability appears analogous to that of BHXRBs suggests that the 
same physical mechanism is at work in generating the variability, regardless
of black hole size.  Furthermore, the fact that PSD break time-scales
are consistent with a linear scaling with black hole mass (assuming the same
accretion state) implies a similar mass-dependent
scaling in the characteristic time-scales
of the underlying process.\cite{utt02,mar03,mch04} \\
The similarities in AGN and BHXRB variability extend beyond the shape of the 
PSD.  Both types of source show a strong linear correlation between the 
rms amplitude of variability and the X-ray 
flux\cite{utt01,vau03a,vau03b,mch04}, which implies
that the variability process is non-linear\cite{utt03b}.
Both BHXRBs\cite{miy89,now99,pot00}
and AGN\cite{pap01,vau03a,mch04}
also show time-scale dependent lags between hard and soft X-ray bands
(with lags of similar magnitude and direction),
and a similar energy-dependence of PSD shape 
above the break frequency, with flatter PSDs at harder energies. 
\cite{now99,nan01,vau03a,mch04} \
These analogies further suggest that clues to the origin of X-ray variability
in AGN can be gained by studying the variability of BHXRBs, for which the 
timing data is more diverse, of higher quality and the phenomenological
understanding of the variability is more advanced. For
example, the various properties
of the rms-flux relation observed in
BH and neutron star XRBs strongly suggest that the variability originates
in the accretion flow itself and is not caused by, e.g. coronal flares 
(although the rate or amplitude of
such flares may be modulated by the accretion flow 
variations).\cite{utt04,gle04} \ Models where the variability is 
due to propagating variations in the accretion flow can also help to 
explain the energy dependent timing properties of 
BHXRBs\cite{kot01} and AGN\cite{vau03a,mch04}. \\
Finally, it is amusing to note that the existence of the PSD breaks observed
in AGN provides indirect evidence for black holes in these objects.  This
is because XRBs which are thought to contain neutron stars do not show
such breaks (their 
PSD slopes remain as $\sim-1$ to high frequencies), but black hole
candidates do\cite{sun00}.  Therefore
by analogy, if we believe that the BHXRBs are aptly named,
we should also believe that black holes power AGN!

\section{Optical variability and the Optical/X-ray relation}
\subsection{A confusing picture}
For many years, most of our knowledge about the variability of radio-quiet 
AGN was learned
from the optical band.  These advances came
 largely thanks to the dedicated efforts of 
teams of observers using ground-based telescopes, primarily to 
monitor the variability of various permitted optical emission lines and their 
response to continuum variations, in an effort to `reverberation map' the
line emitting regions of AGN and so also determine 
their masses.\cite{rob94,pet98} \ A useful
byproduct of these campaigns is a wealth of data on continuum variability
stretching back many years.  Not surprisingly, this variability also appears
to be red-noise, but on short time-scales the amplitude of variability
is much smaller than seen in the X-rays.  The optical continuum
emission in AGN is thought to come primarily from
the thermal emission of the accretion disk,\cite{kor99} \
with longer wavelength emission mainly
originating from larger radii where the disk is cooler.
However, the simultaneous
nature of optical and UV variations (with minimal lag) led
to suggestions that the variable optical/UV emission is driven by
X-ray reprocessing in the disk,\cite{kro91} \
causing the shorter-time-scale X-ray variations to be `washed out' in
the optical by light
travel-time effects.  This mechanism for optical variability can be 
simply tested by searching for correlated optical and X-ray variability.  
Before {\it RXTE}, efforts to search for optical/X-ray correlations were
compromised by the difficulty of obtaining good quality X-ray monitoring to 
match that in the optical, although hints of an optical/X-ray correlation
were observed.\cite{cla92,ede96} \ Interesting
progress was made with short-time-scale
variability, with optical and X-ray monitoring of NGC~4051
showing negligible optical variability during
large-amplitude X-ray variations, ruling out an origin of X-rays and optical 
photons from the same electron population.\cite{don90} \\ 
\begin{wrapfigure}{r}{6.6cm}
\centerline{\includegraphics[width=6cm,height=7cm]{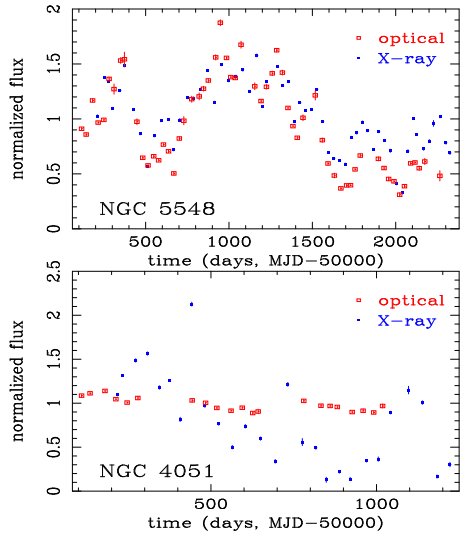}} 
\caption{Comparison of optical (5100$\AA$) and X-ray light curves
of NGC~5548 and NGC~4051.} \label{optvarcomp}
\end{wrapfigure}
The launch of {\it RXTE} allowed much better data to be obtained, and
accordingly
the situation immediately became more confusing.  First, 
month-long {\it IUE} and
{\it RXTE} monitoring of NGC~7469 showed no correlation between continuum 
flux variations in UV and X-ray bands, although the X-ray spectral index 
does appear to correlate with the UV flux, suggestive of Compton cooling by 
the UV photons.\cite{nan98,nan00} \ However, NGC~3516 showed no such 
correlation, with optical and X-ray variations that appear to be
unrelated on both short and long time-scales.\cite{ede00,mao02} \
Intensive optical and X-ray monitoring of NGC~4051, has
finally revealed correlated X-ray and optical variability in this source on
long and short time-scales, although the optical variability
remains surprisingly weak compared to the X-rays (few \% 
fractional rms versus $>50$\%).\cite{pet00,she03,mas02} \ In contrast to these
results, the best optical/X-ray correlation observed in an AGN so 
far is seen in NGC~5548, which shows strong and 
highly correlated variability in both bands on time-scales of months and 
longer.\cite{utt03} \  The difference in optical variability amplitudes
between NGC~5548 and NGC~4051 is highlighted in Fig.~\ref{optvarcomp}.  
Despite both AGN showing strong long-term X-ray variability, only NGC~5548 shows strong optical variability.  In fact, on long time-scales,
the optical variability amplitude of NGC~5548 is even larger than 
the amplitude of X-ray variability, which seems to rule out 
models where optical variability is due solely to reprocessing in the disk. 
This is because in these models the amplitude of
optical variability should be smaller than the amplitude of
X-ray variability, because reprocessed optical emission
would be diluted by the emission from viscous dissipation in the 
reprocessing disk.

\subsection{Explaining the complex optical/X-ray behaviour}
This complicated picture of optical variability in AGN might be explained
if we consider the different origin of the optical emission compared to
the X-ray emission.  The X-rays are presumably produced in optically thin 
material close to the
central black hole, at similar relative radii (i.e. in Schwarzschild radii, $R_{S}$ ) in different
AGN, i.e. irrespective of the black
hole mass (as is implied by the similar X-ray timing behaviour of AGN and 
BHXRBs, despite the huge difference in mass).  However, if the optical
emission originates from optically thick material 
(mainly via viscous dissipation but probably also through reprocessing),
we expect the relative radius of the optical emitting region to depend on
the black hole mass, because disk temperature is thought to scale as 
${\it mass}^{-\frac{1}{4}}$.  In AGN with lower BH mass and higher accretion rate 
(in terms of the Eddington rate), such as NGC~4051, 
most optical emission will originate relatively far from the central 
black hole ($>$few hundred $R_{S}$), due to the relatively
high disk temperature in these objects.  In contrast, higher mass and 
lower accretion rate AGN, such as NGC 5548
 will possess lower temperature disks and so 
their optical emission will originate from closer to the central black hole,
and closer to the X-ray emitting region.  If the inner disk is also less stable
than the outer disk, it is perhaps not suprising that we see strongly variable
optical emission in NGC~5548 which is well correlated with X-rays, but we
see only weakly varying optical emission in NGC~4051.  The situation is
likely to be complicated by the effects of reprocessing and Compton cooling,
so that a combination of these effects could produce the range
of different optical/X-ray relations that are observed.  For example, in
NGC~4051 the {\it XMM-Newton} monitoring suggests that the optical band lags
the X-rays by $\sim0.14$~days\cite{mas02}, whereas longer term monitoring
suggests that on longer time-scales the optical {\it leads}
by 2 days\cite{she03}.  This
intriguing result may suggest that different processes produce the optical
variations on different time-scales.

\end{document}